# Ontology based data warehouses federation management system

Naoual MOUHNI[1], Abderrafiaa EL KALAY[2]

[1] Department of mathematics and computer sciences , University Cadi Ayyad, Faculty of sciences and technologies
Marrakesh, 40000, Morocco
nmouhni@gmail.com

[2] Department of mathematics and computer sciences , University Cadi Ayyad, Faculty of sciences and technologies
Marrakesh, 40000, Morocco
elkalay@hotmail.fr

**Abstract**
Data warehouses are nowadays an important component in every competitive system, it's one of the main components on which business intelligence is based. We can even say that many companies are climbing to the next level and use a set of Data warehouses to provide the complete information or it's generally due to fusion of two or many companies. these Data warehouses can be heterogeneous and geographically separated , this structure is what we call federation, and even if the components are physically separated, they are logically seen as a single component. generally, these items are heterogeneous which make it difficult to create the logical federation schema ,and the execution of user queries a complicated mission. In this paper, we will fill this gap by proposing an extension of an existent algorithm in order to treat different schema types (star , snow flack) including the treatment of hierarchies dimension  using ontology

## 1. Introduction

A Data Warehouse represents the enterprise-wide "single source of truth" and corporate memory of all business process data [3], it is "a subject oriented, non-volatile, integrated, time variant collection of data in support of management's decisions." as defined by Bill Inmon in 1990, the father of data warehouses.

In some cases, one data warehouse is not sufficient to provide a complete information about a fact, which makes grouping multiple data warehouses the only solution. e.g. in the context of a hotel chain that is geographically distributed in many countries, it may have several heterogeneous warehouses to store and analyse data about customers reservations.

this set of warehouses is what we call " a data warehouse federation".

Federated data warehouses are different than distributed Data warehouses, in order that distributed data warehouses can refer different subjects and there is a strict rule in data distribution (horizontal, vertical...) which make it easy to integrate the query results by using join or sum operations [7].

In federated system, the user send his query without having an idea about the location of data or its structure, the set of data warehouses is seen as a whole and the result is the combination of data warehouses components results.

The components in FDWS (Federated data warehouse system) can differ in aspects such as : data model, query language and data semantic [9].

So, a FDWS must contains the following elements [8] :
- An integration procedure of the schemas of the component warehouses giving the logical schema of the federation.
- A query language for user who does not need to know the schemas of the component warehouses.
- A procedure which enables decomposition of user queries to the federation into sub-queries which are sent to the component warehouses

the warehouse federation system management is first based on a logical schema called the federation schema , which integrates all the components schemas. to create this schema, we must integrate all the other local schemas, without loosing information. During this integration, it may be difficult to decide keeping or not an information by using the procedure shows in [8], which compare every measure to the one in the existing federation schema, if it exists  only the location of this measure which is

characterized by the couple $\left(D_0^i, b\text{'\_}name\right)$ is added, where :

$D_0^i$ : represent the fact table in the data warehouse i.

$b\text{'\_}name$ : represent the name of the measure

else, a new measure is added to the schema .

An algorithm is implemented to integrate dimensions attributes, respecting the same logic.

In fact, this algorithm present its limits in case we have a measures or a dimension attributes that refer to the same subject, and represented by two different terms in data warehouses local schemas and it doesn't treat the relationship that could been between attributes and the case of hierarchical dimensions.

Our approach consist of using an application ontology defined in [10] as "a description of knowledge necessary to achieve a particular task and that allows to use the same programming language as the application programming language ", to fill this gap instead of using only Meta data that does not fully represent the semantic relationship between local schema measures and dimension attributes, and those of the federation schema.

In this article, we propose an ontology based data warehouses federation management system to solve the problem of semantic heterogeneity during federation schema creation, based on hotel chain data warehouse sources.

Moreover, in our knowledge, there is no studies that used ontology in a federation context to solve this problem, which justify our choice.

Then in section 2; we present and analysis in summary a set of related works.

## 2. Related works

In all domain research, It is always worth considering the others work , discuss it and check if we can refine and extend it for our particular purpose.

In computer sciences, reusing existing sources is one of the reasons that made the development of this domain possible.

Warehouses federation according to Sheth and Larson [9], and that appears in [7] and [4], is a set of data warehouses that are heterogeneous, autonomous and dispersed. Every component can continue its local operations and at the same time participate in federation.

It's for the better that all the integration operations be done without interrupting the process of component data warehouses.

There are no many studies on the data warehouse federation, however, R. Kern, K. Ryk, and Ngoc Thanh Nguyen, proposed a framework for building logical schema and query decomposition in data warehouse federations [7], they developed an algorithm to integrate component schemas into one global logical federation schema.

But this algorithm presents some limits in order to treat the case of warehouses with star schema only, and it doesn't consider the hierarchical dimensions and all the heterogeneity types, which are described in [9] as the difference in structure, where different data models provides two different structural primitives. then , differences in constraints ,differences in query languages and semantic heterogeneity.

Semantic heterogeneity, is one of the biggest problem that faces information integration nowadays, it occurs when two synonym terms from two different sources describe the same subject [1] ( e.g: schedule and timetable are synonyms but we have to show it to the system) .

one of the solutions to fill this gap is using ontology, which is according to [6] " ontology is a formal explicit description of concepts in a domain of discourse (classes (sometimes called concepts)), properties of each concept describing various features and attributes of the concept (slots (sometimes called roles or properties)), and restrictions on slots (facets (sometimes called role restrictions)). An ontology together with a set of individual instances of classes constitutes a knowledge base. In reality, there is a fine line where the ontology ends and the knowledge base begins."

According to their use, we distinguish many types of ontologies, Generic Ontology, Domain ontology, Application ontology, Representation ontology, The ontology of methods, tasks and resolution of problems, Light ontology and rich ontology[2].

Even if using ontology may resolve the heterogeneity problem in federated data warehouses, it is not yet used in this context, and all the solutions proposed are based on Meta data repositories, which solve the problem of structure definition but not the semantic issues.

## 3. Our contribution

3.1 Presentation of the solution

Our work is an extension to[7] algorithm to create the global logical federation schema .

R. Kern, K. Ryk, and N. Nguyen, proposed an algorithm of integration of component schemas into a federated logical schema. They assume that all warehouses are with star schema, so they do not deal with hierarchies in dimensions.

In fact, even with a star schema the hierarchy for dimension are stored are stored in the dimensional table itself.

Whereas, in a snow flack schema, a dimension table have more or more parent tables, and hierarchies are broken into separate tables in snow flake schema . These hierarchies helps to drill down the data from topmost hierarchies to the lowermost hierarchies[5].

Our objective is an improvement of this integration algorithm to cover heterogeneous schemas (snow flack or star schema). And use ontology as a tool to solve the semantic heterogeneity problem instead of using meta data only.

We propose a federation data warehouse management system (FDWS), which cover :
- Improved algorithm for schemas integration using application ontology
- A query analysis and decomposition tool .
- An ontology-based integration Algorithm for query results.

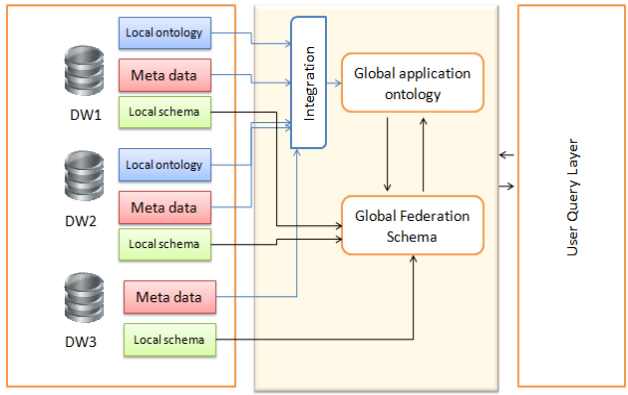

Fig. 1. Structure of the proposed Data warehouses federation management system

1. Every federation component may or not have its own local application ontology, which is written in OWL language describing the semantic of every attribute and measure, and describe the relationship between items and hierarchies of dimensions by using *is_a* and *parentOf* relations.
2. This local ontologies are exported to the logical layer ontology repositroy, besides that a meta data xml file is loaded into the federation system to describe data structure.
3. The user query is analyzed by the FDWS, decomposed, executed on the selected components
4. The query results are integrated using ontology to solve the heterogeneity problem .

3.2 Integration schema's algorithm

In our case, the input can be with different schemas types ( star , snow flack), so to treat the dimension hierarchies we propose the following algorithm:

**Annotation**
We use the same notation as [7].

*Input.*

$P_i^j$ as the set of parents of a dimension defined by $P_i^j = (D_1,...,D_n)$

$H_p$ a Data warehouse schema defined as $H_i = (D_0^i, D_1^i,...,D_{\alpha i}^i)$

$F$ an existing federation schema defined by $F = (D_0, D_1,...,D_m)$

*Output.*

$F$ the federation after integration with $H_p$.

Other notations are used:
$a\_name$ : name of attribute $a$
$b\_name$ : name of measure $b$
$D_x \sim D_y$ : $D_x$ is similar to $D_y$ (based on ontology and meta data OR expert's decision)
$a_x \equiv a_y$ : $a_x$ is similar to $a_y$ (based on ontology and meta data OR expert's decision)
$b_x \Leftrightarrow b_y$ : similar measures (based on ontology and meta data OR expert's decision)

*Recall of the Measure integration algorithm.*
R. Kern, K. Ryk, and N. Nguyen in [7], defined a measure integration algorithm as follow:
For each measure from input data warehouse try to find corresponding measure in federation schema. If such a measure exists in federation schema add a mapping between them. If none of the federation measures corresponds to the current one add it to the federation and make a mapping between new measure and the current one.

**Dimension Integration**
In every iteration of the algorithm, the global schema is being updated by integrating parents of dimensions, then integrating the dimension it self.
1. For each dimension from a component schema, using ontology, we extract the set of this dimension parents, this set can be equal to $\emptyset$ or contains one or many items.
   a. For each parents item, we look for similarity in $F$, if it contains a similar structure, we compare its attributes with the existing one, in case two attributes are similar, we add a new location to the attributes inventory represented by the couple $\left(D_y^p, a\_name\right)$, else, we add the attribute as a new one to the dimension. In case that the attribute doesn't exist in the target dimension, we add a new attribute.
   b. After integrating all the dimension parents, we integrate using the same operations the dimension it self.

$foreach \quad D_y^p \quad \text{in} \, H_p, y = 1, 2, ..., \alpha p$
$\quad if \quad P_y^p \neq \emptyset$
$\quad\quad foreach \quad \dim ension \quad D_i \quad \text{in} \, P_y^p$
$\quad\quad\quad if \quad \exists D_t \in F : D_t \sim D_i$
$\quad\quad\quad\quad foreach \quad attribute \quad a'inD_i$
$\quad if \quad \exists a \in D_t : a \equiv a' \wedge a \quad is \quad characterized$
$by \quad \left(a\_name, list\right)$
$\quad\quad list = list \bigcup \{(D_i, a'\_name)\}$
$\quad else$
$\quad\quad D_t = D_t \bigcup \{(a'\_name, \{(D_i, a'\_name)\})\}$
$\quad endif$
$\quad endforeach$
$else$
$\quad D_t = \emptyset$
$foreach \quad a"in \quad D_i$
$\quad D_t = D_t \bigcup \{(a"\_name, \{(D_i, a"\_name)\})\}$

$\quad endforeach$
$\quad F = F \bigcup \{D_t\}$
$endif$
$if \quad \exists D_t \in F : D_t \sim D_y^p$
$\quad foreach \quad attribute \quad a'inD_y^p$
$\quad\quad if \quad \exists a \in D_t : a \equiv a' \wedge a \quad is \quad characterized$
$by \quad \left(a\_name, list\right)$
$\quad\quad list = list \bigcup \{(D_y^p, a'\_name)\}$
$\quad else$
$\quad\quad D_t = D_t \bigcup \{(a'\_name, \{(D_y^p, a'\_name)\})\}$
$\quad endif$
$endforeach$
$else$
$\quad D_t = \emptyset$
$\quad foreach \quad a'''in \quad D_y^p$

$\quad D_t = D_t \bigcup \{(a'''\_name, \{(D_y^p, a'''\_name)\})\}$
$\quad endforeach$
$\quad F = F \bigcup \{D_t\}$
$endif$
$endforeach$

**4. Example**

We consider that we have two data warehouses which represent the sources of our federation system. The first component is with start schema, so hierarchies dimension are represented in dimension itself. e.g. the hierarchy $Country \rightarrow \text{Re} gion \rightarrow City$.

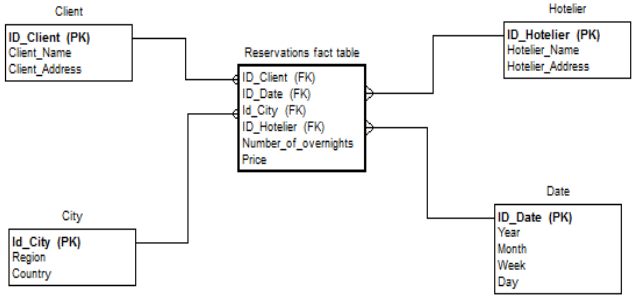

Fig. 2. A star schema of hotel reservations

The second component, is a snow flack schema representing Hotel reservations. this schema contains some hierarchies of dimensions.

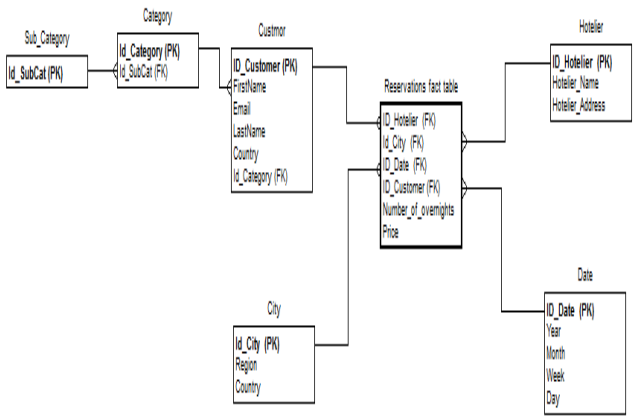

Fig. 3. A snow flack schema for hotel reservations

After applying the proposed integration algorithm we get the global schema as follow:

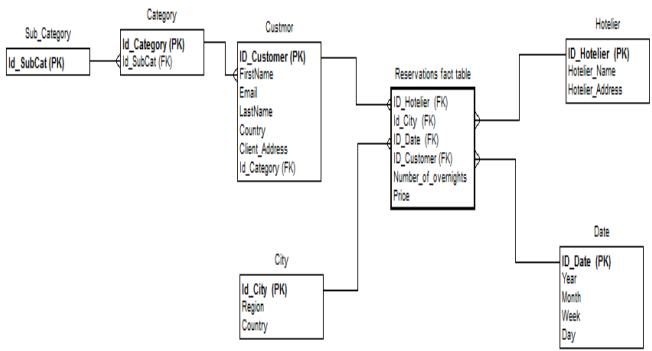

Fig. 4. The result of components schemas integration

Let consider two data ware houses, the first one (**Fig2**) with a star schema and the second one (**Fig3**) is a snow flack schema related to a reservation management in a hotel chain.

1. We first extract ontologies and metadata files from different nodes, in the integration layer of the FDWS, then include new entries into the global ontology repository.
2. Then we integrate fact tables by testing the existence of this table in the global federation schema, if it exists, we compare its measures to the existing ones referring to the ontology repository.
3. next step is to integrate dimensions and hierarchies dimension, e.g: we first integrate the client dimension from DW1 into the global schema, then when we try to include Customer dimension, witch is a synonym of client dimension, so referring to the ontology repository we don't add it as a new dimension, and we compare its attributes with clients attributes.

Based on *parrentof* relationship mentioned in ontology files, between Customer/client and Category and sub_Category we integrate this hierarchy.

## 5. Implementation

The integration schema algorithm was implemented using Java API Jena, to manipulate RDF language from java application. We are using two data warehouses; the first one with a star schema and has an ontology written in OWL/RDF, the second data warehouse with a snow flack schema and has no local ontology.
Metadata files and OWL/RDF files are mapped into xml file and transferred into the network to the Federated data warehouses management system.

## 6. Conclusion

In this article, we have presented a part of our data warehouses federation management system. In particular the process of creating the federation schema based on the integration of local schemas using application ontology. Which makes possible to treat the hierarchies of dimensions by analyzing the parentOf relationships, and make the it easy to automate the integration process in federation context.